\documentclass{llncs}

\usepackage{amsfonts}
\usepackage{amsfonts}
\usepackage{pifont}

\usepackage[noend]{algorithmic} 
\usepackage{algorithm}          

\newcommand{\INPUT}   {\REQUIRE}
\newcommand{\OUTPUT}  {\ENSURE}

\newcommand{\LET}{\textbf{let}}
\renewcommand{\algorithmicrequire}{\textbf{Input:}}
\renewcommand{\algorithmicensure} {\textbf{Output:}}
\renewcommand{\algorithmiccomment}[1]{$/\!/$ #1} 

\newcommand{\equivc}{\;\mathring\equiv\;}

\newcommand{\comment}[1]{}

\title{Craig Interpolation for \\ Quantifier-Free Presburger
Arithmetic\thanks{This research is supported by an award from
IBM Research and by the European Union as part of the FP7-STREP
MOGENTES project.}}

\author{Angelo Brillout\inst{1} \and Daniel Kroening\inst{2} \and Thomas Wahl\inst{1}}
\institute{Computer Systems Institute, ETH Zurich \and Computing Laboratory, Oxford University}
\begin{document}
\spnewtheorem*{soundness}{Soundness}{\itshape}{}
\maketitle


\begin{abstract}
{\em Craig interpolation} has become a versatile algorithmic tool for
improving software verification. Interpolants can, for instance, accelerate
the convergence of fixpoint computations for infinite-state systems. They
also help improve the refinement of iteratively computed lazy abstractions.
Efficient interpolation procedures have been presented only for a few
theories. In this paper, we introduce a complete interpolation method for
the full range of {\em quantifier-free Presburger arithmetic} formulas. We
propose a novel convex variable projection for integer inequalities and a
technique to combine them with equalities. The derivation of the
interpolant has complexity low-degree polynomial in the size of the
refutation proof and is typically fast in practice.

\end{abstract}

\section{Introduction}\label{intro}

A {\em Craig interpolant}, or simply \emph{interpolant}, for an
inconsistent pair of formulas $A$ and $B$ is a formula $I$ that is implied
by $A$, inconsistent with $B$, and contains only variables occurring in
both $A$ and $B$ \cite{DBLP:journals/jsyml/Craig57a}. In other words, a
Craig interpolant is weaker than $A$, but still strong enough to be
inconsistent with $B$, and therefore provides an ``explanation'' of the
inconsistency in terms of the common variables. In his original theorem,
Craig showed that an interpolant exists for any two inconsistent
first-order formulas $A$ and $B$.

Craig interpolants have proven to be useful in many areas. McMillan
suggested to use them in an over-approximating image
operator~\cite{DBLP:conf/cav/McMillan03}, which has led to a considerable
advance in SAT-based model checking. For infinite-state systems,
interpolants can significantly improve the refinement step in lazy predicate
abstraction~\cite{DBLP:conf/cav/McMillan06}. Methods to efficiently compute
interpolants are known for propositional logic and linear arithmetic over
the reals with uninterpreted
functions~\cite{DBLP:journals/jsyml/Pudlak97,DBLP:journals/tcs/McMillan05}.
For these theories, an interpolant can be derived in linear time from a
deductive proof of inconsistency of $A$ and $B$.

\emph{Presburger arithmetic} is a popular theory for modeling computer
systems, for example to describe the behavior of infinite-state
programs~\cite{DBLP:conf/cav/BultanGP97}. It was shown to be decidable by
quantifier elimination \cite{entscheidbarkeit_PA}, which is, however, of
double-exponential complexity. Fortunately, formulas arising in system
specification and verification are mainly
quantifier-free~\cite{DBLP:journals/lmcs/SeshiaB05,DBLP:conf/fmcad/GaneshBD02}.
In this paper we therefore focus on \emph{quantifier-free Presburger
arithmetic} (QFP). An interpolant between two inconsistent QFP formulas $A$
and $B$ can be computed by existentially quantifying the variables that
occur only in $A$, followed by quantifier elimination. This approach is,
however, prohibitively expensive.

\paragraph{Contribution}

In this paper, we propose an algorithm that extracts an interpolant
directly from a proof of inconsistency of $A$ and $B$. Our algorithm
extends the framework of Pugh's
\emph{Omega test}~\cite{DBLP:conf/sc/Pught91}. We present suitable
deduction steps in the form of inference rules. Following a suggestion by
McMillan~\cite{DBLP:journals/tcs/McMillan05}, the rules are augmented with
{\em partial interpolants}---expressions that are transformed step by step
to yield an interpolant of the initial formulas $A$ and $B$ once a
contradiction has been reached. We present our algorithm for conjunctions
of equalities and inequalities; interpolants for an arbitrary
Boolean skeleton can be obtained using the framework described in
\cite{DBLP:journals/tcs/McMillan05}.

For conjunctions of equalities, our algorithm exploits the fact that
\emph{exact} variable projection is efficient for certain fragments of
QFP. We therefore treat equalities separately in the first part of the
paper and describe such a projection procedure. Our procedure supports
\emph{stride constraints}, i.e., quantified equalities expressing
divisibility relationships. For conjunctions of inequalities, we show that
deriving an interpolant requires the \emph{strongest convex} projection
(which may be inexact) and give an efficient algorithm for computing this
projection. Finally, we describe the first interpolation method that
combines conjunctions of integer equality and inequality constraints.

\paragraph{Related work}

For propositional logic, several interpolation methods have been
presented~\cite{DBLP:journals/jsyml/Pudlak97,DBLP:conf/cav/McMillan03,DBLP:journals/jsyml/Krajicek97}.
In addition to the work by McMillan \cite{DBLP:journals/tcs/McMillan05},
Rybalchenko et al.\ propose an algorithm for linear arithmetic over the
reals with uninterpreted functions that circumvents the need for an
explicit proof~\cite{DBLP:conf/vmcai/RybalchenkoS07}. For integer
arithmetic, McMillan considers the logic of difference-bound
constraints~\cite{DBLP:conf/cav/McMillan06}. This logic, a fragment of QFP,
is decidable by reducing it to arithmetic over the reals. Difference-bound
constraints are, however, not sufficient to express many typical program
constructs, such as integer
divisibility~\cite{DBLP:journals/lmcs/SeshiaB05}.

For interpolating SMT (satisfiability modulo theory) solvers, which involve
calls to theory-specific provers, combination frameworks have been
presented in~\cite{cimatti08,DBLP:conf/cade/YorshM05}. In~\cite{beyer08},
an SMT solver is used to derive interpolants for rational linear arithmetic
with uninterpreted functions. In~\cite{jcg2008}, separate interpolation
procedures for two theories are presented, namely (i) QFP restricted to
conjunctions of integer linear (dis)equalities and (ii) QFP restricted to
conjunctions of stride constraints. The combination of both fragments with
integer linear inequalities is, however, not supported. Our work closes
this gap, as it permits predicates involving all types of constraints. Such
predicates arise naturally for instance in inductive invariant discovery,
as argued in \cite{jcg2008}.


\paragraph{Outline}
This paper is organized as follows.
%
Section~\ref{prelem} contains background and terminology. In
section~\ref{eq_str}, we present the rules for computing interpolants of
inconsistent equality and stride constraints. Section~\ref{ineq} does the
same for inequality constraints and for combinations of both. In
section~\ref{complex}, we discuss the time complexity of our algorithm.

\section{Preliminaries}\label{prelem}

\subsection{Craig Interpolants}

Two QFP formulas are {\em inconsistent} if their conjunction is
unsatisfiable. We define $\mathcal{V}(\phi)$ to be the set of variables
occurring in a (quantifier-free) formula $\phi$. For any two formulas $A$
and $B$, we write $\mathcal{L}_A$ for the set of variables {\em local}
to~$A$, i.e., $\mathcal{L}_A = \mathcal{V}(A)\setminus\mathcal{V}(B)$.
Analogously, we write $\mathcal{G}$ for the {\em global} (common) variables
of $A$ and $B$, i.e., $\mathcal{G} = \mathcal{V}(A)\cap\mathcal{V}(B)$. The
quantifier-free formulas $A$ and $B$ are {\em equisatisfiable}, denoted $A
\equivc B$, if existentially quantifying their respective local variables
produces two logically equivalent formulas, i.e., $\exists \mathcal{L}_A. A
\equiv \exists \mathcal{L}_B. B$. Let $\bot$ and $\top$ represent the
Boolean values \emph{false} and \emph{true}, respectively.
\begin{definition}\label{craig_inter}

A {\em (Craig) Interpolant} for two inconsistent quantifier-free formulas
$(A, B)$ is a formula $I$ such~that:

\begin{description}
	\item[\textit{(1)}] $ A \models I$,
	\item[\textit{(2)}] $ (B,I) \models \bot$, and
	\item[\textit{(3)}] $\mathcal{V}(I)\subseteq \mathcal{G}$.
\end{description}	
\end{definition}%
As an example, let $A$ and $B$ be the (inconsistent) formulas $x=y+1 \land
z=y$ and $x=y$, respectively. An example of an interpolant $I$ for $A$ and
$B$ is $x=y+1$.

\subsection{Quantifier-free Presburger Arithmetic}

Presburger arithmetic is the first-order theory defined by the structure
$\langle \mathbb{Z},=,\le,$ $+\rangle$, i.e., quantified linear {\em
integer} arithmetic with arbitrary Boolean connectives. In 1929,
M.~Presburger presented a quantifier elimination procedure for this logic,
which gives rise to a decision procedure~\cite{entscheidbarkeit_PA}.

We consider in this paper {\em quantifier-free} Presburger arithmetic with
stride predicates, denoted QFP. Atoms, henceforth called {\em constraints},
are of the form
\begin{displaymath}
t\bowtie0 \ \ (\bowtie \; \in \{=,\le\}) \ \ \ \mbox{or} \ \ \ d \mid t \ \ (d \in \mathbb{N}_{\ge 2}) \,,
\end{displaymath}
where $t$ is a term of the form $\sum_{j\in J}a_{j} x_{j} + c$. We call
these atoms {\em equality}, {\em inequality} and {\em stride} constraints.

The stride predicates $d \mid t$ specify divisibility properties of a term
$t$, e.g., $2\mid x$ denotes that $x$ is even. We refer to $d$ as the {\em
periodicity} of a stride constraint. To motivate the need for stride
predicates, consider the equalities $x-2y=0$ and $x-2z-1=0$, whose only
quantifier-free interpolant is $2\mid
x$~\cite{DBLP:journals/tcs/McMillan05}.

We say that two constraints $\sum_{j\in J}a_{j} x_{j} + c \bowtie 0$ and
$\sum_{j\in J}b_{j} x_{j} + d\bowtie 0$ are {\em parallel} if for every
$j\in J$, $a_{j} = b_{j}$ or for every $j\in J$, $a_{j} = -b_{j}$.
A \emph{unit coefficient} is a coefficient $a_j$ with $|a_j|=1$.

QFP formulas are constructed using the usual Boolean connectives. We adopt
the method in \cite{DBLP:journals/tcs/McMillan05} to reduce reasoning over
arbitrary Boolean combinations of constraints to reasoning over
conjunctions.
Despite the stipulation of being quantifier-free, we permit a restricted
form of quantification in QFP, namely over {\em finite} sets of
integers. Formulas containing such quantifications are semantically
quantifier-free since they can be rewritten using a finite disjunction.

\subsection{Equisatisfiability-Preserving Manipulations}

\subsubsection*{Tightening of inequalities}

Let $g:=gcd(\{|a_j|: j \in J\})$ be the greatest common divisor of the
coefficients in the term $t = \sum_{j\in J}a_{j} x_{j} + c$ of an
inequality $t \le 0$. We say the inequality is {\em tight} if $g$ divides
$c$. Every inequality can be transformed into an equivalent tight form by
replacing $c$ with $g\lceil\frac{c}{g}\rceil$. We refer to $\mathcal{T}(f)$
as the tight form of an inequality $f$. (Note that an equality constraint
$t = 0$ is unsatisfiable if $g$ does not divide $c$.)


\subsubsection*{Homogenization}

Let $Q(x)$ be a formula over $x$. We \emph{homogenize} $Q(x)$ by computing
an equisatisfiable formula $F(\sigma)$ over a new variable $\sigma$ (but
without $x$) such that all coefficients of $\sigma$ are unit
coefficients. This is achieved as follows:
\begin{enumerate}

\item Compute the least common multiple $l := lcm\{|a| : \; a$ is a
coefficient of $x$ in some constraint$\}$.

\item Multiply each constraint over a term containing a multiple $ax$ of
$x$ by $\frac{l}{|a|}$; for a stride constraint $d \mid t$ this means to
multiply both $d$ and $t$ by $\frac{l}{|a|}$. The result is a formula
$Q'(x)$ equivalent to $Q(x)$ where all coefficients of $x$ are either $l$
or $-l$.

\item Replace every occurrence of $lx$ in $Q'(x)$ with a new variable
$\sigma$ and conjoin the result with the new constraint $l \mid \sigma$.

\end{enumerate}

The obtained formula $F(\sigma)$ and the original $Q(x)$ are
equisatisfiable, with $\sigma$ having unit coefficients everywhere, as shown
by Cooper~\cite{cooper}. A formula is called {\em $\sigma$-homogenized} if
all occurrences of $\sigma$ have unit coefficients.

\subsubsection*{Exact projection}

We define a projection method that is based on~\cite{cooper}; our method is
simpler since it assumes an $x$-homogenized conjunction $Q(x)$ of constraints containing at most one inequality. {\em Exact projection} amounts to eliminating $x$
from $Q(x)$, resulting in an equisatisfiable formula. We distinguish
two~cases:

\begin{itemize}

\item If there is at least one equality containing $x$ in $Q(x)$, let $eq$
be any such equality. Since every occurrence of $x$ has a unit coefficient,
$eq$ can be rewritten as $x=t$. Now obtain a new, equisatisfiable formula
$Q'(t)$ by dropping the conjunct $eq$ from $Q(x)$ and replacing $x$ by $t$
everywhere else.

\item Otherwise, let $l:=lcm\{d: d$ is a periodicity of some
stride constraint containing $x\}$. Remove any inequality over $x$ from $Q(x)$ resulting in a $Q'(x)$. Eliminate $x$ by replacing $Q'(x)$ with $\exists
i\in\{0,\ldots,l\}.Q'(i)$. The result is equisatisfiable to $Q(x)$.

\end{itemize}

We denote by $proj(Q(x),x)$ a procedure that first $x$-homogenizes $Q(x)$ and
then returns an equisatisfiable formula by exact projection. We extend this
procedure to act on a formula $Q$ and a \emph{set} of variables $V$, denoted
$proj(Q,V)$, by applying $proj$ to $(Q(x),x)$ for all $x\in V$ in any
order.

\section{Equality and Stride Constraints}\label{eq_str}

In this section, we present an algorithm for deriving an interpolant for
two inconsistent formulas $A$ and $B$ that are conjunctions of stride and
equality constraints. The algorithm is based on an elimination procedure
for equality and stride constraints (section \ref{elim_eq}). The procedure
is refined in section \ref{proj_eq} by annotating its steps with partial
interpolants.

\subsection{Eliminating Equality and Stride Constraints}\label{elim_eq}

\newcommand\hatmod{\,\widehat{\rm{mod}}\,}

We use an algorithm proposed by Pugh~\cite{DBLP:conf/sc/Pught91} for
eliminating the equalities from the system of constraints. For this purpose, we need
a slightly modified ``centered'' modulus function $\!\hatmod\!$, defined as
$a\hatmod b:= a-b\lfloor\frac{a}{b}+\frac{1}{2}\rfloor$.
We write $t \hatmod b$ to denote $\sum_{i\in J} (a_i\hatmod b)x_i +
(c\hatmod b)$ for a term $t$ of the form $\sum_{i\in J}a_ix_i+c$. This follows
from distributivity of $\hatmod\!$.

The elimination algorithm first replaces each stride constraint $d \!\mid\! 
t$ by the equisatisfiable equality $d\sigma +t =0$, where $\sigma$ is a
fresh variable. What remains is a system of equalities. Consider the
following equality involving variable $x$:
\begin{eqnarray}
a x + t = 0\,. \label{eq}
\end{eqnarray}
If $x$ has a unit coefficient, we can eliminate the equality by deleting it
from the system and replacing every occurrence of $x$ by~$-at$. Otherwise,
by applying the $\!\hatmod\!$ operator to both sides of equality (\ref{eq})
and introducing a fresh variable $\sigma$, we obtain the new constraint
\begin{eqnarray}
\left(a \hatmod  m \right)x + \left( t \hatmod m \right) = m \sigma \label{unit_eq}
\end{eqnarray}
where $m=|a|+1$. Since $a \hatmod m = -sign(a)$, variable $x$
in~(\ref{unit_eq}) has a unit coefficient. Thus, we can eliminate $x$
in~(\ref{eq}) and in all other constraints involving~$x$. As shown
in~\cite{DBLP:conf/sc/Pught91}, the absolute values of the coefficients in
the new equality resulting from~(\ref{eq}) have decreased, eventually
resulting in an equality with a unit coefficient. This equality can be
eliminated without applying the $\hatmod$ operator.

We call the original constraint~(\ref{eq}), which is used to derive a
constraint with a unit coefficient, the {\em pivot equality}, denoted
$eq_p$. Let $\phi$ be a conjunction of equalities. We denote by
$elim(\phi)$ the procedure that eliminates all equalities in $\phi$ using
pivot equalities $eq_p$ chosen according to some heuristics -- we refer the
reader to~\cite{DBLP:conf/sc/Pught91} for such a heuristic. Note that each
elimination of an equality leaves the remaining system equisatisfiable to
the original one. Therefore, if the procedure ever encounters an
unsatisfiable equality, it immediately returns $\bot$, indicating
inconsistency of the original constraints. Otherwise, the original system
is eventually reduced to an equality of the form $c=c$ for some
constant~$c$; the procedure returns $\top$. Note that, since we assume $A$
and $B$ to be inconsistent, $elim$ never returns $\top$ unless we consider
combinations of equalities and inequalities and the inconsistency is due to
the inequalities (section~\ref{comb_eq_ineq}).

\subsection{Interpolation for Equality and Stride Constraints}\label{proj_eq}

The first part of our contribution follows. We introduce rules in order to
derive an interpolant from a proof of inconsistency of the linear equality
formulas $A$ and $B$. To do so, we borrow the notion of a {\em partial
interpolant} from~\cite{DBLP:journals/tcs/McMillan05}.

\begin{definition}\label{eq_inter}
A {\em partial equality interpolant for $(A,B)$} is a conjunction of linear
equalities $\phi^A$ such that:
\begin{description}
	\item[\textit{(1)}] $A \models \phi^A$, and
	\item[\textit{(2)}] $(B,\phi^A) \models \phi$, and
	\item[\textit{(3)}] if $\phi$ contains an unsatisfiable equality, then $\mathcal{V}(\phi^A) \subseteq \mathcal{G}$.
\end{description}
where $A$, $B$ and $\phi$ are conjunctions of equalities. We write $(A,B) \vdash \phi \; [\phi^A]$ if we can
\emph{derive} the partial interpolant $\phi^A$ from $(A,B)$.
\end{definition}
%
Observe that if $\phi\equiv\bot$, definitions~\ref{craig_inter}
and~\ref{eq_inter} coincide, with $\phi^A$ as the interpolant.

Consider now a proof of inconsistency of the two conjunctions $A$ and $B$
of equalities. The proof consists of a sequence of proof rule
applications. We extend these rules to apply to partial interpolants that
are attached to antecedent and consequent of the rules. The partial
interpolants are transformed to eventually result in an interpolant for
$(A,B)$. We first present a rule to introduce hypotheses and the
corresponding partial interpolant for $(A,B)$ in the proof tree.
\begin{eqnarray*}
\mbox{\sc{HypEq}}
\frac{}
		 		 {(A,B) \vdash A \wedge B \; [A]}
\end{eqnarray*}
The partial interpolant is simply $A$. Note that {\sc HypEq} introduces all
equalities simultaneously. The soundness proof for this rule, showing that
the derived partial equality interpolant conforms to the three conditions
of definition \ref{eq_inter}, is straightforward.

The next rule eliminates the equality constraints as mentioned in section
\ref{elim_eq}. The rule results in a partial interpolant where $A$ is
projected by elimination of the variables local to $A$:
\begin{eqnarray*}
\mbox{\sc{ElimEq}}
\frac{
  \begin{array}{lrl}
    (A,B) \vdash & \hspace*{24pt} A\wedge B \hspace*{4pt}&[A]\hspace*{45pt}
  \end{array}}{
  \begin{array}{llcrrl}
    (A,B) \vdash & elim(A\wedge B) &[proj(A,\mathcal{L}_A)]
  \end{array}}
\end{eqnarray*}
If function $elim(A \wedge B)$ returns $\bot$, the (final) interpolant is
$proj(A,\mathcal{L}_A)$. Note that, in this interpolant, every
variable local to $A$ has been eliminated by $proj$ and that no new
variable has been introduced.

\begin{soundness}[of \sc{ElimEq}]
To show the soundness of the rule, we argue that the rule \emph{preserves}
the three conditions of definition \ref{eq_inter}. Regarding the first
condition, the fact that $A \models proj(A,\mathcal{L}_A)$ follows
immediately from the soundness of Cooper's projection procedure. Since
$A\wedge B\equivc B\wedge proj(A,\mathcal{L}_A)\equivc elim(A\wedge B)$, we
know that $(B, proj(A,\mathcal{L}_A)) \models elim(A\wedge B)$. This shows
condition 2. The $proj$ procedure eliminates every local variable to $A$
and thus $\mathcal{V}(proj(A,\mathcal{L}_A))\subseteq \mathcal{G}$. This
shows condition 3.
\qed
\end{soundness}

\begin{example}
We would like to find an interpolant for $A \equiv (6 \mid 3z-2y-2)$ and $B
\equiv (6x-y=0)$. Using the {\sc HypEq} rule, we introduce both constraints
and the partial interpolant. We apply the {\sc ElimEq} rule to the result:
\begin{eqnarray*}
\mbox{\sc{ElimEq}}
\frac{\displaystyle (A,B) \vdash 6\sigma+3z-2y-2=0 \wedge 6x-y=0 \; [6 \mid 3z-2y-2]}
		 		 {(A,B) \vdash 6\sigma-12x-2=0 \; [\exists i \in \{0\ldots6\}.\;(6 \mid i-2y-2)\wedge (3\mid i)]}
\end{eqnarray*}
We eliminate $y$ by applying $elim$, since $y$ has a unit coefficient in
$6x-y=0$. However, the substitution of $6x$ for $y$ produces a
contradiction since $gcd(6,12)$ does not divide $2$. We project the partial
interpolant by eliminating the only local variable $z\in \mathcal{L}_A$. To
do so, $proj$ z-homogenizes the partial interpolant, resulting in $(6 \mid
\sigma-2y-2)\wedge (3\mid\sigma)$ and finally in the interpolant $\exists i
\in \{0,\ldots,6\}.\:(6 \mid i-2y-2)\wedge (3\mid i)$.  \qed
\end{example}

\section{Inequality Constraints}\label{ineq}

This section presents a method for deriving an interpolant for two
inconsistent formulas $A$ and $B$ that are conjunctions of inequalities. We
first review the variable elimination procedure used in the Omega test
(Section~\ref{FM}). We then introduce the notion of strongest convex
projection (Section~\ref{str_conv_proj}), which is necessary to refine the
procedure with partial interpolants (Section~\ref{inter_ineq}).

\subsection{Fourier-Motzkin variable elimination for QFP}\label{FM}

W.\ Pugh adapted the \emph{Fourier-Motzkin} (FM) variable elimination
method to QFP~\cite{DBLP:conf/sc/Pught91}. This section briefly reviews
this method. In the following, $t_1$ and $t_2$ are two terms not containing
the variable $x$, and $a,b$ are positive integers. Consider the two
inequalities
\begin{eqnarray} \label{up_low}
a x + t_1 \le 0 \;\; \; \mbox{ and } \; -b x + t_2 \le 0 \;.
\end{eqnarray}
These inequalities are upper (left constraint) and lower (right
constraint) bounds on $x$. Equivalently, we get
\begin{eqnarray}
a t_2 \le a b x  \le - b t_1 \label{befor_proj}
\end{eqnarray}
by multiplying the upper and lower bounds by $b$ and $a$, respectively. The
FM method eliminates variable $x$ by deducing the following inequality
from~(\ref{befor_proj}):
\begin{eqnarray}
\mathcal{T}(a t_2 + b t_1 \le 0)\; \label{after_proj} 
\end{eqnarray}
where $\mathcal{T}(a t_2 + b t_1 \le 0)$ denotes the tight form of $a t_2 +
b t_1 \le 0$. Inequality~(\ref{after_proj}) is a projection of
(\ref{befor_proj}) that eliminates $x$. Note that (\ref{befor_proj})
implies (\ref{after_proj}), but not generally vice versa: the two are not
equisatisfiable. We therefore speak of an \emph{inexact} projection. If the
distance between the upper and the lower bound is less than $ab$, there may
or may not be a solution to the following equation:
\begin{eqnarray}
\mathcal{T}(- a b +1 \le a t_2 + b t_1  \le 0)\; . \label{thin_area}
\end{eqnarray}
Note that in~(\ref{thin_area}), the strict inequality $-ab < at_2 + bt_1$
has been replaced by the equivalent inequality $-ab + 1 \le at_2 +
bt_1$. In geometrical terms,~(\ref{thin_area}) describes the ``thin'' part
of the polyhedron~(\ref{befor_proj}). If no inconsistency is found by
(inexact) projection of all inequalities, i.e., only inequalities of the
form $-p \le0$, $p\in \mathbb{N}_{\ge 1}$ remain, one must check for
solutions in this ``thin'' part.

For this purpose, Pugh introduced {\em splinters}. Given are the
bounds~(\ref{up_low}) leading to inexact projection. An equality $-b x + t_2
+ i = 0$ is added to the original set of inequalities.  This equality is
eliminated as explained in section~\ref{elim_eq} and the FM algorithm is
called recursively. This is done for each $i\in \{0,\ldots,s\}$ where $s=\lfloor (|nb|
- |n|-b)/|n|\rfloor$ and $n$ is the negative coefficient of $x$ with the largest absolute value in any
inequality. If all splinters of all inexact projections produce an
inconsistency, then the original system of inequalities is unsatisfiable. We
refer the reader to~\cite{DBLP:conf/sc/Pught91} for further details.


\subsection{Strongest Convex Projection}\label{str_conv_proj}

Consider the case that an inconsistency is reached without the need for
splinters, i.e., inexact projection is sufficient to show inconsistency of
$(A,B)$. Since the inexact projection (\ref{after_proj}), being a single
inequality, describes a \emph{convex} region, there is also a convex
interpolant. In order to compute it, we introduce the notion of {\em
strongest convex} projection, i.e., the {\em strongest} projection
expressible with {\em one} inequality.
Formally, we introduce:
\begin{definition}\label{def_str_conv_proj}
For lower and upper bounds $ax+t_1 \le 0$ and $-bx+t_2 \le 0$, let $t' \le
0$ be the tight form of $at_2+bt_1 \le 0$, and let $m \geq 0$. Inequality
$t' + m \le 0 $ is the {\em strongest convex projection} of these bounds if
there is no integer $i$ such that:
\begin{displaymath}
(a t_2 \le a b x  \le - b t_1) \models (t' + i \le 0) \models (t' + m \le 0)\,.
\end{displaymath}
\end{definition}

We now present a new method to compute the strongest convex projection of a
lower and an upper bound; see algorithm
\ref{strongest_convex_projection}. The bounds are converted into the
inequality $-ab + 1 \le at_2 + bt_1 \le 0$. Tightening this inequality
results in a constraint of the form $-c' \le t' \le 0$, which can
equivalently be expressed as the quantifier-free formula $\exists i \in
\{-c',\ldots,0\}.\:t' = i$. This is our pivot equality (line~1). This
equality, conjoined with the lower and upper bounds, can be checked for
satisfiability, thus revealing which integers $i$ are feasible in the
``thin'' part of the polyhedron (\ref{befor_proj}). We perform this check
in line 2 using an elimination procedure modified from section
\ref{elim_eq}: we find an equality with a unit coefficient, rewrite it into
the form $y = t_y$ and replace every occurrence of $y$ in the inequality
and in $eq_p$ by $t_y$. This is repeated until the pivot equality $eq_p$ is
reduced to $\top$, resulting in the bounds of the form as given by $f_1$
and $f_2$ in line 2. Note that the constants $c$ and $d$ depend on $i$.
\begin{algorithm}[htbp] 
  \begin{algorithmic}[1]
    \INPUT {lower bound $ax + t_1 \le 0$, upper bound $-bx + t_2 \le 0$}
    \OUTPUT {strongest convex projection of these bounds}
    \STMT {\LET\ $eq_p \ = \ (\exists i \in \{-c',\ldots, 0\}.\:t'=i)$} \COMMENT {tight form of $-ab + 1 \le at_2 + bt_1 \le 0$}
    \STMT {\LET\ $f_1 = (t'_1 + c(i) \le 0)$ and $f_2 = (t'_2 + d(i) \le
      0)$ be the tight inequalities resulting from reducing $eq_p$ to $\top$}
    \IF {$t'_1 = t'_2$ or $f_1$ and $f_2$ are not parallel}
      \RETURN {$t' \le 0$}
    \ELSE[$t'_1 = -t'_2$]
      \STMT {\LET\ $A = \{i: -c' \le i \le 0 \ \wedge \ c(i)+ d(i)\le 0\}$}
      \IF {$A \not= \emptyset$}
        \RETURN {$t' - (\min A) \le 0$}
      \ELSE
        \RETURN {$t'-c'+1 \le 0$}
      \ENDIF
    \ENDIF
  \end{algorithmic}
  \caption{Strongest convex projection}
  \label{strongest_convex_projection}
\end{algorithm}

We demonstrate some aspects of algorithm \ref{strongest_convex_projection}
with the following example.
\begin{example}
Suppose the following bounds are given:
\begin{eqnarray}
x + 3y - 2\le 0 \ \wedge \ x-3y+1 \le 0\;. \label{example1}
\end{eqnarray}
In line 1, the algorithm tightens the ``thin'' part of the projection which
is $-8 \le 6x-3 \le 0$. The result is $6x=0$, i.e., here $c'=0$ and
$i=0$. In line 2, this equality is substituted into (\ref{example1});
tightening produces two inequalities $3y\le 0$ and $-3y+3\le 0$. These are
parallel with unequal terms (case $t'_1=-t'_2$, line 5). Since
$A=\emptyset$, the strongest convex projection $6x+1\le0$ is returned in
line 10.\qed
\end{example}

In the following section, we continue example 2 and demonstrate why the
notion of strongest convex projection is necessary for deriving partial
interpolants.

\subsection{Interpolation for Inequality Constraints}\label{inter_ineq}

The notion of partial interpolants for inequalities is defined as follows.
\begin{definition}\label{def_in}
A {\em partial inequality interpolant for $(A,B)$} is an inequality $t^A \le 0$ such that:
\begin{description}
	\item[\textit{(1)}] $A \models t^A \le 0$,
	\item[\textit{(2)}] $B \models t - t^A \le 0$, and
	\item[\textit{(3)}] $\mathcal{V}(t^A \le 0) \subseteq \mathcal{V}(A)$ and $\mathcal{V}(t - t^A)\subseteq \mathcal{V}(B)$.
\end{description}
where $A$, $B$ are conjunctions of inequalities and $t$, $t^A$ terms. We
write $(A,B) \vdash t \le 0 \; [t^A \le 0]$ if we can \emph{derive} the
partial interpolant $t^A \le 0$ from~$(A,B)$.
\end{definition}
Observe that if $t$ is a positive constant, $t - t^A \le 0$ is a
contradiction and $t^A \le 0$ is an interpolant for $(A,B)$.

We now present rules that implement the FM elimination procedure for
QFP. As with equalities, these rules also compute partial interpolants that
preserve the properties of definition~\ref{def_in}. When introducing a
hypothesis in the proof, the partial interpolant depends on the origin of
the hypotheses:
\begin{displaymath}
\mbox{\sc{HypIn}}
\frac{}{(A,B) \vdash t \le 0 \; [\chi_A(t \le 0)]} \;
(t \le 0) \in (A,B)
\end{displaymath}
where $\chi_A(t \le 0)$ is defined to be $t \le 0$ if $t \le 0\in A$ and
$0\le 0$ otherwise.

The next rule projects inequalities. When combing two inequalities to
achieve projection, the same linear combination is applied to the partial
interpolants.
\begin{eqnarray*}
\mbox{\sc{Proj}}
\frac{
	\begin{array}{lrl}
	  (A,B) \vdash \hspace*{12pt} &  a x + t_1 \le 0 &\hspace*{1pt} \; [t_1^{A} \le 0] \\
		(A,B) \vdash &                -b x + t_2 \le 0 &\hspace*{1pt} \; [t_2^{A} \le 0]
	\end{array} 
\hspace*{69pt}
}
{(A,B) \vdash \mathcal{T}(a t_2 + b t_1 \le 0) \; [\mathcal{T}(at_2^A + bt_1^A + m \le 0)]}
{ a,b \in \mathbb{N}_{\ge 1} }
\end{eqnarray*}
where $m$ is a constant such that if inexact projection occurs,
$\mathcal{T}(t_2^A + t_1^A + m \le 0)$ is the strongest convex projection,
and otherwise $m=0$.

\begin{soundness}[of \sc{Proj}]
We check if the conditions of definition~\ref{def_in} are preserved.
Condition 1 is straightforward. From the premises, we know that $B
\models ax + t_1 - t_1^A\le 0$ and $B \models -bx + t_2 -
t_2^A\le 0$ and thus $B\models (a t_2 + b t_1)-(a
t_2^A + b t_1^A) \le 0$, which is convex. Tightening a constraint only
increases the constant $c$ of the corresponding inequality and since the
projection (if any) of the partial interpolant is the {\em strongest}, we
conclude $B\models \mathcal{T}(a t_2 + b
t_1)-\mathcal{T}(a t_2^A + b t_1^A +m)$. This proves condition 2. The fact
that $\mathcal{T}$ does not change the coefficients in a constraint
emphasizes the similarity with the linear arithmetic method described in
\cite{DBLP:journals/tcs/McMillan05}. As in this work, projection by
eliminating a variable $x$ also eliminates $x$ in the partial interpolant.
Since every variable has to be eliminated to obtain an inconsistency, the
interpolant does not contain any local variable. This shows condition 3.
\qed
\end{soundness}

\begin{example}
We show how to derive an interpolant for $A\equiv x+3y-2\le0\wedge
x-3y+1\le0$ and $B\equiv -x\le0$. We write $t\le 0 \; [t^{A} \le 0]$ instead
of $(A,B) \vdash t \le 0 \; [ t^{A} \le 0]$ and we do not show how to
introduce hypotheses to save space. First, we project the two inequalities
from $A$ by eliminating $x$:
\begin{eqnarray*}
\mbox{\sc{Proj}} \
\frac{
  \begin{array}{rcll}
        -x & \le & 0 & [0 \le 0] \\
    x+3y-2 & \le & 0 & [x+3y-2 \le 0]
  \end{array}}{
  \hspace*{35pt}
  \begin{array}{rcll}
    3y & \le & 0 & [x+3y-2\le 0]
  \end{array}}
\hspace{3mm}
\mbox{\sc{Proj}} \
\frac{
  \begin{array}{rcll}
    -x & \le & 0 & [0 \le 0] \\
   x-3y+1 & \le & 0 & [x-3y+1 \le 0]
  \end{array}}{
  \hspace*{10pt}
  \begin{array}{rcll}
    -3y+1 & \le & 0 & [x-3y+1\le 0]
  \end{array}}
\end{eqnarray*}
We can now derive a contradiction by eliminating $y$:
\begin{eqnarray*}
\mbox{\sc{Proj}} \
\frac{
  \begin{array}{rcll}
     3y   &\le & 0 & [x+3y+1 \le 0] \\
    -3y+1 &\le & 0 & [x-3y-2 \le 0]
	\end{array}}{
	\hspace*{13pt}
  \begin{array}{rcll}
	  3 &\le &0 & [6x+1\le 0]
	\end{array}}
\end{eqnarray*}
Note that the interpolant $6x+1\le0$ is the strongest convex projection of
$x+3y+1\le0$ and $x-3y-2 \le 0$, which was computed in the example at the
end of section \ref{str_conv_proj}. Observe that the standard projection of
the partial interpolant according to equation~(\ref{after_proj}) is
$6x\le0$, which does not interpolate $(A,B)$.\qed
\end{example}

\subsubsection*{Splinters}

If the FM procedure for QFP introduces splinters as described in
section~\ref{FM}, the Omega test is called recursively for each
splinter. More precisely, in case of an inexact projection when applying
the {\sc Proj} rule, the interpolation algorithm is called upon each pair
$(A,B)_i$ and $(A,B)_{\geq s+1}$ defined as $(A\wedge t_2^A+i=0, B\wedge
t_2-t_2^A=0)$ and $(A\wedge t_2^A+s+1\le0,B\wedge t_2-t_2^A\le0)$,
respectively.

%

If all splinters produce an inconsistency, i.e., all pairs $(A,B)_i$ and
$(A,B)_{\geq s+1}$ are inconsistent, the original system is
unsatisfiable. We can construct an interpolant for $(A,B)$ from the
respective interpolants $I_i$ and $I_{\geq s+1}$ for $(A,B)_i$ and
$(A,B)_{\geq s+1}$ as follows:
\begin{eqnarray*}
\hspace{2cm}
\mbox{\sc{Splin}}
\frac{
\begin{array}{ll}
(A,B)_{\geq s+1}& \vdash \bot \; [I_{\geq s+1}] \hspace*{31pt}\\
(A,B)_i& \vdash \bot \; [I_i]
\end{array}
}
{
\begin{array}{lllll}
(A,B)\hspace*{22pt} &\vdash \bot \; [\vee_i^s \; I_i \vee I_{\geq s+1}]
\end{array}
}
\begin{array}{l}
\mbox{ for all }i\in\{0,\ldots,s\}
\\
\\
\end{array}
\end{eqnarray*}

\begin{soundness}[of {\sc Splin}]
We show that the derived interpolant conforms to
definition~\ref{craig_inter}. We denote by $A_i$, $B_i$, $A_{\ge s+1}$ and
$B_{\ge s+1}$ the respective components of the pairs $(A,B)_i$ and
$(A,B)_{\ge s+1}$. Condition 1 follows from $A_i\models I_i$ and $\vee_i^s
A_i \vee A_{\ge s+1}\equiv A$ follows condition 1, i.e., $A\models \vee_i
A_i \vee A_{\ge s+1}$. Condition 2 follows from the unsatisfiability of
$\vee_i^s A_i\wedge B_i \wedge A_{\ge s+1}\wedge B_{\ge s+1}$ and $\vee_i
B_i \vee B_{\ge s+1}\equiv B$. Finally, since $\mathcal{V}(A_i) =
\mathcal{V}(A)$ and $\mathcal{V}(A_{\ge s+1})=\mathcal{V}(A)$ condition 3
also holds.
\end{soundness}

\begin{example}
In example 3 we first eliminate $x$ and then $y$ by using {\sc Proj}. Note
that if we reverse this order, no inconsistency is reached by using {\sc
Proj} only. This is due to the inexact projection of $x + 3y - 2\le0$ and
$x - 3y + 1\le 0$ by elimination of $y$. In this case, the number of
splinters that must be derived is given by $s=(3*3-3-3)/3=1$.

The Omega test is then called recursively for each pair $(A,B)_i$,
$i\in\{0,1\}$ and $(A,B)_{\ge 2}$ given by $(A\wedge x-3y+1+i=0,B)$ and
$(A\wedge x-3y+1+3\le 0,B)$, respectively. The pairs $(A,B)_i$ contain both
inequalities and equalities. In the next section, we show how to derive an
interpolant for $(A,B)_1$. Once each pair $(A,B)_i$ and $(A,B)_{\ge2}$ has
been proved inconsistent, we combine their respective interpolants with the
{\sc Splin} rule:
\begin{eqnarray*}
\mbox{\sc{Splin}} \
\frac{
  \begin{array}{lrl}
    (A,B)_0 &\vdash \bot & [2x\le0\wedge 3\mid (x+1)] \\
    (A,B)_1 &\vdash \bot & [2x\le0\wedge 3\mid (x+2)] \\
    (A,B)_{\ge2}& \vdash \bot & [6x+1 \le 0]    
  \end{array}}{
  \begin{array}{lrl}
    (A,B) &\vdash \bot & [(6x+1 \le 0)\vee(2x\le0\wedge 3\mid x+2)\vee(2x\le0\wedge 3\mid x+1)]
  \end{array}}
\end{eqnarray*}
Note that the result is indeed an interpolant for $(A,B)$.\qed

%
\end{example}

\section{Putting it all together}\label{comb_eq_ineq}

\subsection{Combining equality and stride constraints with inequalities}

We now turn to the most challenging part of this work, namely deriving an
interpolant for two inconsistent formulas $A$ and $B$ that are conjunctions
of equality, inequality and stride constraints. In this section, we denote
by $E_A$ and $E_B$ the conjunction of equality and stride constraints of
$A$ and $B$, respectively. In order to detect any inconsistency, the Omega
test begins by eliminating stride and equality constraints from the system,
i.e., the {\sc HypEq} and {\sc ElimEq} rules are applied to the pair
$(E_A,E_B)$. We distinguish two cases:
\begin{enumerate}

\item[(i)] Suppose an unsatisfiable equality was found during the
elimination. In this case, $E_I:=proj(E_A,\mathcal{L}(E_A))$ is an
interpolant for $(E_A,E_B)$. From the validity of $A \models E_A$, $B
\models E_B$ and $\mathcal{L}(E_A) \subseteq \mathcal{L}(A)$, it
follows that $E_I$ also an interpolant for $(A,B)$. That is, we derive an
interpolant for $(A,B)$ using the {\sc HypEq} and {\sc ElimEq} rules only,
\emph{without} considering the inequalities at all.

\item[(ii)] Otherwise, all equality and stride constraints are successfully
eliminated. In this case, for each $x=t_u$ derived with the $\hatmod$
operator, the Omega test replaces each occurrence of $x$ in {\em every}
constraint of $A$ and $B$, not only in the equalities. Eventually, a new
pair $(A',B')$ consisting only of inequalities remains. The formula
$A'\wedge B'$ is then equisatisifibale to $A\wedge B$.

\end{enumerate}
To formalize the second case, we denote by $\phi\{x\leftarrow t_u\}$ the
result of substituting the term $t_u$ for every occurrence of variable $x$
in $\phi$. By $\phi \{\vec{x}\leftarrow \vec{t_u}\}$ we denote the
\emph{sequence} of substitutions performed, in this order, during the
equality elimination process. The formulas $A'$ and $B'$ are then given by
$A\{\vec{x}\leftarrow \vec{t_u} \}$ and $B\{\vec{x}\leftarrow \vec{t_u}
\}$, respectively.

%
%

We can now derive new partial interpolants for $(A',B')$ using the {\sc HypIn}, {\sc
Proj} and {\sc Splin} rules, with $A'$ and $B'$ in place of $A$ and
$B$. Once a contradiction is reached, the obtained interpolant will be
valid for $(A',B')$, but not for $(A,B)$. More precisely, since the terms
$\vec{t}_u$ may contain new variables, the generated interpolant may also
contain a variable not occurring in $(A,B)$. The problem is to map an interpolant for $(A',B')$ to an
interpolant for $(A,B)$.

We address this problem as follows. Let $t^{A'}\le0$ be an interpolant for
$(A',B')$. We show below how to compute a partial interpolant
$t^A\le0$ for $(A,B)$ such that $t^{A'}=t^A\{\vec{x}\leftarrow
\vec{t}_u\}$. We then demonstrate that $proj(t^A\le0\wedge
E_A,\mathcal{L}_A)$ is an interpolant for $(A,B)$. This is formalized using
the following rule:
\begin{eqnarray*}
\mbox{\sc{Comb}} \
\frac{ (A',B')  \vdash \bot \; [t^{A'}\le 0]\hspace*{67pt}} 
		 		 {(A\hspace*{3pt},B\hspace*{3pt}) \vdash  \bot \; [proj(t^A\le0\wedge E_A,\mathcal{L}_A)]}
\ \
\begin{array}{ll}
  t^{A'} = t^A \, \{\vec{x}\leftarrow \vec{t_u}\}, \\
  (A,B) \vdash t\le 0 \, [t^A \le 0]
\end{array}
\end{eqnarray*}
The partial interpolant $t^A\le 0$ that is needed to apply this rule is
computed by ``postponing'' the substitutions. That is, after applying the
{\sc Proj} rule, the partial interpolant is kept in the form
$at^A_1+bt^A_2\{\vec{x}\leftarrow \vec{t}_u\}\le 0$ instead of
$at^A_1\{\vec{x}\leftarrow~\vec{t_u}\} + bt^A_2\{\vec{x}\leftarrow
\vec{t}_u\}\le0$.
\begin{soundness}[of \sc{Comb}]
We show that the derived interpolant satisfies
definition~\ref{craig_inter}. First, we observe that applying a
substitution before a projection is equivalent to applying it after the
projection, i.e., $a t_1\{\vec{x}\leftarrow\vec{t_u}\} +
bt_2\{\vec{x}\leftarrow \vec{t}_u\}\le0
\equiv at_1+bt_2\{\vec{x}\leftarrow \vec{t}_u\}\le 0$, where the
substitutions are naturally extended to terms. Thus, there is no immediate
need to apply the substitutions $\{\vec{x}\leftarrow\vec{t}_u\}$ before
projecting using the {\sc Proj} rule. More precisely, we can always derive
a partial interpolant such that $(A',B')\vdash t\le0 \{\vec{x}\leftarrow
\vec{t}\}[t^A\le0\{\vec{x}\leftarrow \vec{t}\}]$. Subsequently, we know
that $t_A\le0$ and $t\le0 $ are linear combinations of inequalities in $A$
and that, if projection occurs, $t^A\le 0$ is the strongest convex
projection. Thus, $t_A\le0$ is a partial interpolant for $(A,B)$.

If an inconsistency is reached, we have derived a partial interpolant for
$(A,B)$ such that $t \{\vec{x}\leftarrow \vec{t}\}=c$ for some positive
constant $c$. Since $t^A \le 0$ is a partial interpolant we conclude
$A\models proj(t^A\le0\wedge E_A)$, which proves condition 1. To prove
condition 2, we first note that $B\wedge proj(t^A\le 0 \wedge E_A)$ and
$(B\wedge t^A\le 0 \wedge E_A) \{\vec{x}\leftarrow \vec{t} \}$ are
equisatisifiable. We know $B\models t-t^A\le 0$ and, thus, conclude
$(B\wedge t\le 0 \wedge E_A) \{\vec{x}\leftarrow \vec{t} \}$. This
contradicts $t \{\vec{x}\leftarrow \vec{t}\}=c$, proving condition
2. Condition 3 follows since $proj$ eliminates all variables local to
$A$. \qed
\end{soundness}

\begin{example}
Consider the pair $(A,B)_1$ given by $(A\wedge x-3y+2=0,B)$, where $A$ and
$B$ are from example 3. There is only one equality to eliminate. Since it
has a unit coefficient, the only substitution is $\{x \leftarrow
(3y-2)\}$. The two partial interpolants resulting in an inconsistency are:
\begin{eqnarray*}
\mbox{\sc{Proj}}\
\frac{
  \begin{array}{lcrclcrrl}
    (A,B)_1 &\vdash & 6y & \le & 0 & [&x-3y+1 &\le 0 &\{x\leftarrow(3y-2)\}] \\
    (A,B)_1 &\vdash & -3y +2 & \le & 0 & [&0  &\le 0 &\{x\leftarrow(3y-2)\}]
  \end{array}}{
  \begin{array}{llcrrl}
    (A,B)_1 &\vdash \hspace*{38pt}\bot \hspace*{8pt} & [&x-3y+1 &\le 0 &\{x\leftarrow(3y-2)\}]
  \end{array}}
\end{eqnarray*}
Note that the subsitutions were not applied to the partial interpolant in
order to determine the final interpolant with the {\sc Comb} rule:
\begin{eqnarray*}
\mbox{\sc{Comb}}\
\frac{
  \begin{array}{llcrrl}
    (A,B)_1 &\vdash \bot & [&x-3y+1 &\le 0 &\{x\leftarrow(3y-2)\}]
  \end{array}}{
  \begin{array}{lrl}
    (A,B)_1 &\vdash \bot & [2x\le 0\wedge 3 \mid(x+2)]\hspace*{52pt}
  \end{array}}
\end{eqnarray*}
The resulting interpolant has been obtained by applying $proj$ to $x-3y+1 \le0\wedge x-3y+2=0$.\qed

\end{example}

\begin{figure}[t]
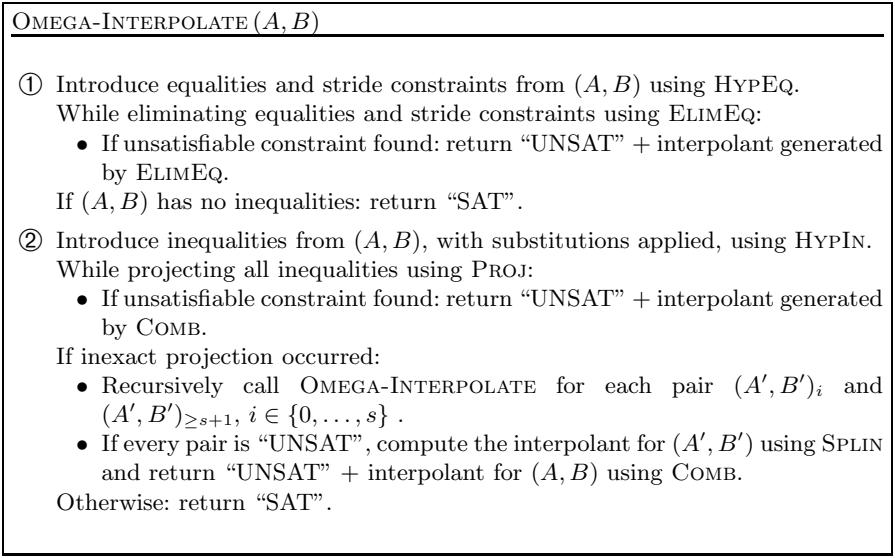

\centering
\framebox{%
\begin{minipage}{.95\textwidth}{$\mbox{\sc Omega-Interpolate}\,(A, B)$}
\hrule
\vspace{1ex}
\begin{itemize}
\item[{\raisebox{-1pt}{\large \ding{192}}}]
Introduce equalities and stride constraints from $(A,B)$ using {\sc HypEq}. \\
While eliminating equalities and stride constraints using \textsc{ElimEq}:
\begin{itemize}
  \item If unsatisfiable constraint found: return ``UNSAT'' + interpolant generated by \textsc{ElimEq}.
\end{itemize}
If $(A,B)$ has no inequalities: return ``SAT''.

\vspace{1ex}

\item[{\raisebox{-1pt}{\large \ding{193}}}]
Introduce inequalities from $(A,B)$, with substitutions applied, using {\sc HypIn}.\\
While projecting all inequalities using \textsc{Proj}:
\begin{itemize}
\item If unsatisfiable constraint found: return ``UNSAT'' + interpolant generated by \textsc{Comb}.
\end{itemize}
If inexact projection occurred:
\begin{itemize}

  \item Recursively call $\mbox{\sc Omega-Interpolate}$ for each pair
  $(A',B')_i$ and $(A',B')_{\ge s+1}$, $i\in\{0,\ldots,s\}$ .

  \item If every pair is ``UNSAT'', compute the interpolant for $(A',B')$
  using {\sc Splin} and return ``UNSAT'' + interpolant for $(A,B)$ using
  \textsc{Comb}.
\end{itemize}

Otherwise: return ``SAT''.
\end{itemize}
\vspace{1ex}
\end{minipage}}
\vspace{1ex}
\caption{The Omega-Test with Interpolation\label{fig:overview}}
\end{figure}


\paragraph{Summary} Fig.~\ref{fig:overview} shows the Omega test extended
by our deduction rules in order to construct (partial) interpolants. Since
the Omega test is complete for conjunctions of equalities, inequalities and
stride constraints and we provide a deduction rule for each of its steps,
the extended algorithm is complete as well.

In practice, we decouple the search for an inconsistency from the
computation of an interpolant. That is, our implementation of {\sc
Omega-Interpolate} takes $A$, $B$ and an inconsistency proof as input and
annotates this proof with partial interpolants. This allows many
optimizations. Substitutions are not performed if all equalities
encountered during step \ding{192} are satisfiable, and no partial
interpolant will be computed for projections that do not lead to an
inconsistency. Throughout the algorithm, arithmetic normalizations, such as
replacing $3y\le 0$ by $y \le 0$, prevent the coefficients from growing
unnecessarily.

\subsection{Time Complexity}\label{complex}

We discuss the worst-case time complexity of our interpolation
algorithm. Let $a$ be the maximum absolute value of any coefficient and any
periodicity occurring in the partial interpolants across the entire
proof. In the original set of constraints, let $w$ denote the maximum
number of variables per constraint in $A$ and $e$ the number of equality
and stride constraints in $A$. The cost of eliminating one variable $x$
using $proj$ is $O(e(w+\log^2 a))$. Let $v$ be the number of local
variables that occur in equalities. Procedure $proj$ is applied $v$ times
in order to eliminate all local variables. In the worst case, the number of
constraints $e$ increases by one after each projection due to
homogenization. In case of an inconsistency in the equalities, the
worst-case time complexity of deriving an interpolant is therefore $O((w +
\log^2 a)(ve + v^2))$.

For inequalities, the {\sc Proj} rule, including computing the strongest
convex projection, has a complexity of $O(w \log a)$. If {\sc Proj} is
applied $p$ times, the overall interpolation complexity is therefore $O(pw
\log a+(w+\log^2 a) (ve+v^2))$. We observed the run-time to be much
smaller in practice, owing to many unit or small coefficients in the
original pair $(A,B)$ (also confirmed
by~\cite{DBLP:journals/lmcs/SeshiaB05}).


%

\section{Conclusion}\label{concl}

We have presented an interpolation method for \emph{quantifier-free
Presburger arithmetic} (QFP). Our method first eliminates equalities and
stride constraints from the system and then projects inequalities using an
extension of the Fourier-Motzkin variable elimination. These steps are
formalized as proof rules that, as a side effect, transform \emph{partial
interpolants} to full interpolants for the given system of constraints. Our
method is the first to enable efficient interpolation for quantifier-free
linear integer arithmetic. In contrast to previous work, it permits
combinations of equalities, inequalities and divisibility properties.

The results presented in this paper are expected to improve model checking
based on counterexample-guided abstraction refinement (CEGAR). As shown in
\cite{jcg2008}, program verification often requires computing inductive
invariants involving constraints over integers. If a candidate invariant
fails, interpolation can aid the discovery of new candidates. Our work
permits the computation of interpolants for formulas given as combinations
of the above-mentioned constraints.

A preliminary implementation of our algorithm shows that for QFP formulas
occurring in practice, the run-time of the algorithm is much better than
the estimated worst-case performance. We contribute this efficiency to
small variable coefficients and a small number of variables per constraint.

\bibliographystyle{splncs}
\bibliography{paper}

\end{document}